\documentclass[aps,prb,preprint,groupedaddress,dvips]{revtex4}
\usepackage[dvips]{graphicx}
\begin{document}
\preprint{}
\title{Localization and Capacitance Fluctuations in Disordered Au Nano-junctions}
\author{M. Bowman, A. Anaya, A. L. Korotkov, and D. Davidovi\'c}
\affiliation{Georgia Institute of Technology, Atlanta, GA 30332}
\date{\today}
\begin{abstract}

Nano-junctions, containing atomic-scale gold contacts between
strongly disordered leads, exhibit different transport properties
at room temperature and at low temperature. At room temperature,
the nano-junctions exhibit conductance quantization effects. At
low temperatures, the contacts exhibit Coulomb-Blockade. We show
that the differences between the room-temperature and low
temperature properties arise from the localization of electronic
states in the leads. The charging energy and capacitance of the
nano-junctions exhibit strong fluctuations with applied magnetic
field at low temperature, as predicted theoretically.

\end{abstract}
\pacs{73.23.-b,73.63.-b,73.21.-b}
\maketitle

\section{Introduction\label{intro}}

The prospect of molecular electronics as a potential alternative
to conventional silicon-based electronics has lead to an increased
interest in fabrication of atomic scale gaps and atomic-scale
contacts between metallic electrodes. Examples include
atomic-scale gaps formed by mechanically controlled break
junctions,~\cite{muller,agrait}
electrodeposition,~\cite{morpurgo,li,boussaad,yu,yu1} and
electromigration~\cite{park,park1,anaya}. In these fabrication
techniques, one can determine whether a junction has atomic-scale
dimensions by changing the conductance of the junction around the
conductance quantum $G_Q=e^2/h$. Discrete steps in conductance of
order $G_Q$ indicate that the contacts have atomic scale
dimensions. This scheme works remarkably well in cases where the
gaps and the contacts are formed in ultra-high-vacuum (UHV)
conditions, such as mechanically controlled break junctions at
cryogenic temperatures.~\cite{agrait}

Some schemes for generating atomic-scale gaps involve exposure of
these gaps to non-UHV environment, such as air~\cite{park,anaya}
or ionic solutions~\cite{morpurgo,li,boussaad,yu,yu1}. In this
case, intermixing between atoms in the leads and impurity
molecules (such as $H_2O$) can degrade the quality of the gaps.
Understanding of electrical conduction in such disordered
atomic-scale gaps and atomic-scale contacts is still lacking.

Recently, Yu and Natelson have studied Au nano-junctions formed by
electroplating from an aqueous solution.~\cite{yu,yu1} Transport
measurements were carried out at both room temperature and
cryogenic temperatures. The authors found different transport
properties at room and low temperature. At room temperature, as
the gap size between two Au leads is reduced by electroplating,
conductance increased in discrete steps of order $G_Q$, suggesting
that the contacts were atomic scale, consistent with the prior
work.~\cite{morpurgo,li,boussaad} In addition, the nano-junctions
were Ohmic at room temperature.

At $T=1.8K$, however, Au junctions with room temperature
conductance $G(300K)\sim G_Q$, the conductance at zero bias
voltage and $T=1.8K$ was suppressed by $\approx 100\%$, which was
referred to as the zero-bias anomaly (ZBA). They argued that ZBAs
displayed a suppression of the density of states in the leads at
the Fermi level, as a result of disorder introduced by the
electroplating process. The disorder was attributed to the grain
boundaries and adsorption of impurities from the solution.

We have reported similar observations in Au nano-junctions formed
by an electric-field-induced migration process.~\cite{anaya} At
room temperature, as the conductance of the junctions increased
from a value below the conductance quantum to above the
conductance quantum, the conductance displayed discrete steps in
conductance, of order $G_Q$. In addition, the room temperature I-V
curves of the samples were linear (Ohmic).

At low temperatures, we found strong ZBAs in samples with
$G(300K)\sim G_Q$, similar to the ZBAs in electroplated Au
nano-junctions. However, samples with $G(300K)<G_Q$ were found to
exhibit Coulomb blockade, proved by the quasiperiodic gate-voltage
dependence of the conductance at $T=0.015K$. Coulomb blockade was
attributed to single-electron charging effects on one or a few
grains in the leads. The data fit exceptionally well the theories
of Coulomb Blockade in the weak~\cite{averin,averin2} and the
strong coupling regimes~\cite{golubev}.

\begin{figure}
\includegraphics[width=0.75\textwidth]{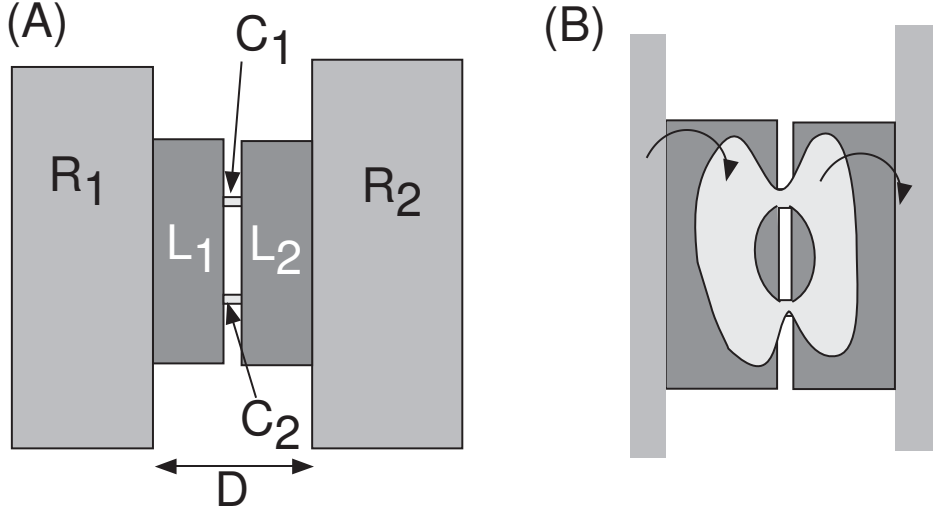}
\caption{A. Disordered Au nano-junction. B. Sequential electron
tunnelling through the nano-junction, via a localized puddle of
electrons.~\label{model}}
\end{figure}

In this paper, we first show that Coulomb blockade in Au
nano-junctions is not restricted to single electron charging on
one or few metallic grains. In fact, Coulomb blockade is observed
when the resistance of the leads is comparable with the resistance
of the contacts, even if there are no apparent grains in the
leads. We propose a general model of a disordered Au nano-junction
containing atomic-scale contacts, which is sketched in
Fig.~\ref{model}. Reservoirs $R_1$ and $R_2$ are bulk Au films,
which are good metals. $C_1$ and $C_2$ are atomic-scale metallic
contacts that are responsible for conductance quantization at room
temperature. $L_1$ and $L_2$ are highly disordered leads, with
room temperature resistances smaller than or comparable to the
resistance of atomic-scale contacts.

The model reconciles the difference between room and low
temperature transport properties of Au nano-junctions, as follows.
The resistivity of the leads is assumed high enough to cause
strong localization. However, the characteristic temperature at
which localization suppresses conductivity in the leads is assumed
to be smaller than $300K$. In this case, the resistance of the
contacts dominates at room temperature, explaining conductance
quantization and Ohmic properties. At low temperatures, however,
the resistance of the leads becomes much larger than the
resistance of the atomic scale contacts, explaining
Coulomb-Blockade and ZBAs. This interpretation of ZBAs in terms of
localization is different from the alternative interpretation in
terms of suppression of the density of states in the
leads.~\cite{yu,yu1} In sec.~\ref{electro} we explain the
difference in more detail.

After our model is presented, we discuss capacitance fluctuations
of the nano-junctions with applied magnetic field. The capacitance
fluctuations in coherent conductors in the charging regime have
been predicted theoretically,~\cite{nazarov} but have not yet been
demonstrated experimentally. The strong disorder combined with the
small size of our nano-junctions makes it possible to study
charging effects in the phase coherent regime, permitting us to
demonstrate and explore capacitance fluctuations.

The paper is organized as follows: In section~\ref{fabrication} we
give a detailed summary of the nano-junction fabrication process
and arrive at the nano-junction model shown in Fig.~\ref{model}-A.
In section~\ref{cb} we present Coulomb-blockade measurements and
discuss electron localization in the leads. In section~\ref{cf} we
discuss capacitance fluctuations. In section~\ref{electro} we
explain the differences between our samples and electroplated
nano-junctions.

\section{Fabrication of gold nano-junctions\label{fabrication}}

The fabrication of Au nano-junctions used in this paper has been
described in Refs.~\cite{anaya,korotkov}. In this section we
summarize the fabrication process. We present new data and new
images of the nano-junctions, which have improved our
understanding of nano-junction properties since prior
publications.

\begin{figure}
\includegraphics[width=0.75\textwidth]{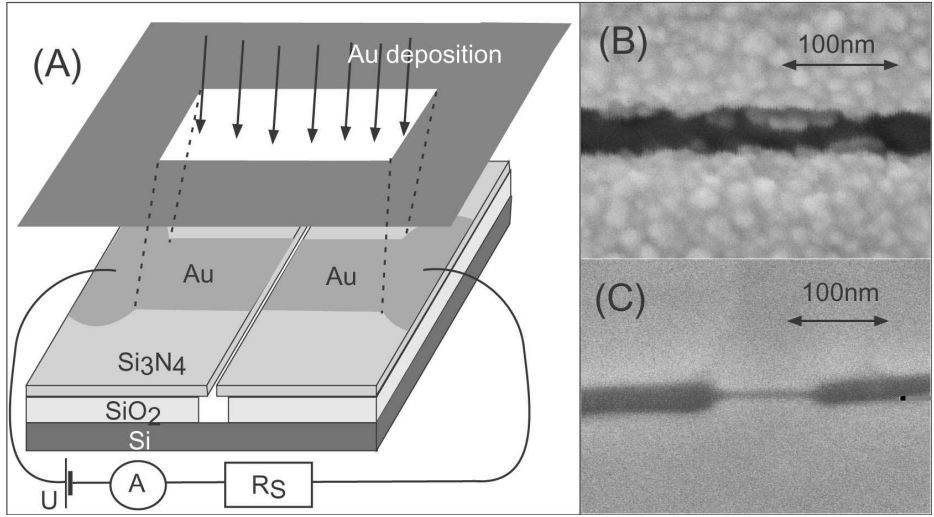}
\caption{A. Deposition of Au over a 70nm wide slit. B. Image of
the gap between two 70 nm thick Au films grown at low bias
voltage. C. Image of the gap between two 70 nm thick Au films
grown at high bias voltage.~\label{fab}}
\end{figure}

To create a nano-junction between two Au films, we deposit Au
atoms over a 70nm wide slit, as shown in Fig.~\ref{fab}-A.  The
slit is created in $Si_3N_4$ using electron-beam lithography and
etching.~\cite{anaya} The large undercut serves to prevent the
connection between two Au films. The exposed length of the slit is
$0.1mm$. The current between the films is monitored in situ. The
current limiting resistor $R_S$ is added in series with the
sample.

Gold deposition is done by thermal evaporation and the deposition
rate is 2.5$\AA /s$. The background pressure of the deposition
chamber measured near the gate valve of the pump is $\sim 10^{-7}$
Torr. Because water molecules outgas from the mask and other
nearby surfaces, the sample pressure is higher. The pressure
measured with a gauge placed near the sample is in the $10^{-6}$
Torr range.

During the deposition, the gap between two gold films is reduced
in proportion to the film thickness. If the bias voltage is weak
($<0.1V$), then the two gold films electrically connect when the
thickness of the film reaches about 70nm. Fig.~\ref{fab}-B shows
the shape of the gap between two Au films of thickness 70nm grown
at $U=0.1V$. The films are not connected in this sample. The edge
of the film is quite rough because of grains sticking to the edge
of the gap. At 70nm thickness, there is a $\sim$50\% chance that
there is a pair of grains attached on the opposite sides of the
gap and that are in electric contact. By stoping the deposition at
the moment when the desired current is detected, we create an
atomic scale gap or an atomic scale contact.

\subsection{Electric field induced surface diffusion}

The bias voltage has a strong influence on the shape and electric
properties of the nano-junctions. In general, polarization effects
from the applied electric field can induce atom migration
processes with a "hierarchy of activation energies".~\cite{mayer}
These processes include electric-field induced surface diffusion,
migration due to localized heating, elastic and plastic
deformation, and field desorption. The activation energy of these
processes depends on both the electric field and the electric
field gradient. It has been demonstrated that surface atom
diffusion caused by the field gradient has the lowest activation
energy.~\cite{mayer}

In our samples, if the voltage applied between the films is large
($\sim 10V$), a strong electric field inside the gap can pull a
pair of protrusions from the opposing sides of the gap.
Fig.~\ref{fab}-C shows the shape of the gap between two Au films
grown at 20V. The edges of the films are much smoother than those
in Fig.~\ref{fab}-B. In addition, the film, in the vicinity of the
gap in Fig.~\ref{fab}-C, is also much smoother than the film in
Fig.~\ref{fab}-B.

These differences can be explained by field induced surface
diffusion. At large bias voltage, roughness along the film edges
(Fig.~\ref{fab}-B) induces field gradients, decreasing the
activation energy for surface diffusion. In response, surface Au
atoms diffuse where the electric field gradient is the strongest,
thereby reducing surface roughness.

In the sample in Fig.~\ref{fab}-C, there is neither mechanical nor
electric contact between the two films. This shows that the
protrusion stopped growing on its own, before a contact could have
established. The two protrusions are almost mirror images of each
other.

Processes such as elastic and plastic deformation and field
desorption are driven by the magnitude of the electric field, not
the field gradient, and therefore can not be responsible for
protrusion growth.
 The electric field in Fig.~\ref{fab}-C is
strongest where the gap is smallest, which would increase the
speed of the protrusion growth due to elastic or plastic
deformation. On the other hand, the electric field gradient is
weak inside this region, thereby decreasing the speed of
protrusion growth due to surface diffusion.


\subsection{Tunnelling Contacts}

\begin{figure}
\includegraphics[width=0.75\textwidth]{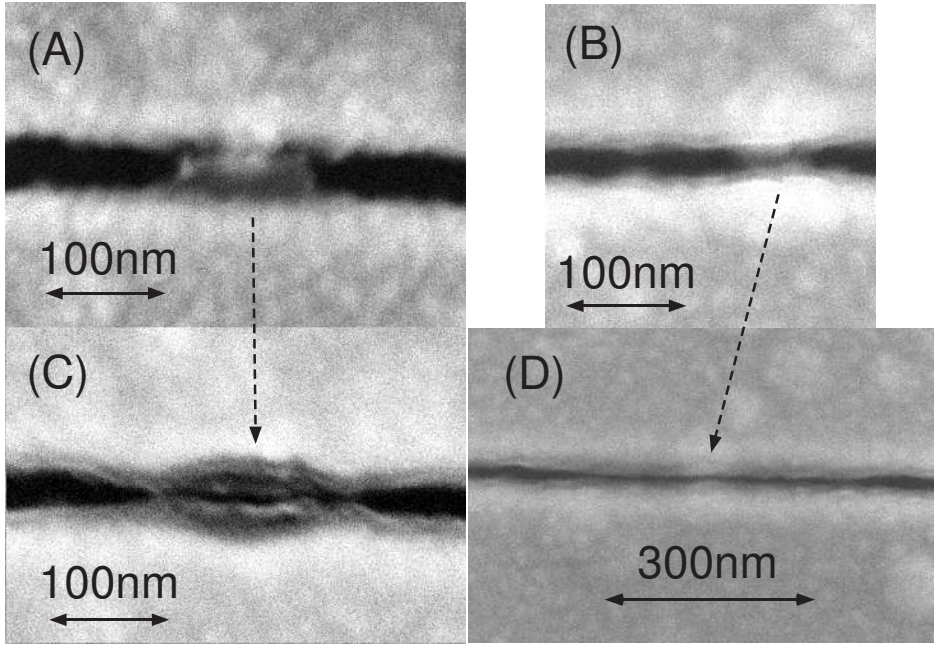}
\caption{A and B: Two tunnelling junctions formed at 10V bias
voltage C and D: The same contacts, after the conductance is
increased above $2e^2/h$.~\label{junctions}}
\end{figure}

For most of our samples, the electric field induced surface
diffusion leads to a contact between the two protrusions.
Fig.~\ref{junctions}-A and B show two such contacts, formed during
growth at 10V. Deposition of Au was stopped as soon as the
slightest electric contact was detected. The electric contact was
exposed to 10V for $\approx 1$s, and then the bias voltage was
quickly reduced  to zero (at a rate of 1V in 10ms). I-V curves
were obtained by measuring current while bias voltage was reduced
to zero. The samples were subsequently transferred to the scanning
electron microscope (SEM) and images were taken.

\begin{figure}
\includegraphics[width=0.75\textwidth]{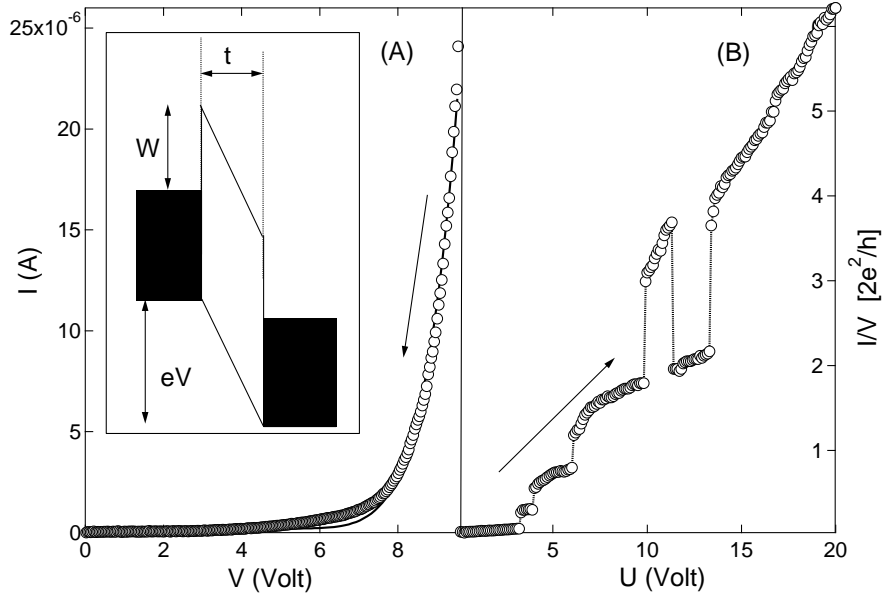}
\caption{A: Circles: I-V curve of a Au nano-junction. Line: fit to
the field emission model. Inset: Schematic of the field emission
model. $W$ is the tunnelling barrier height and $t$ is the barrier
thickness. The best fit parameters: $W=5.8eV$ and $t=10.1\AA$. B:
Discrete steps in conductance of order $2e^2/h$ in a current
limited Au nano-junction.~\label{quantized}}
\end{figure}

The resistance of the junctions is large compared to the
resistance quantum. The I-V curves fit quite well the model of
field emission through a tunnelling barrier with a barrier height
close to the work function of Au (5.1 eV) and the barrier
thickness of about $10\AA$, as shown in Fig.~\ref{quantized}-A.
The fitting is described in Ref.~\cite{korotkov}

The key point that we want to make here is that the voltage drop
of 10V is not distributed uniformly through the leads. It is
localized within a single tunnelling junction. If it were
otherwise, the I-V curve would exhibit less barrier bending than
that in Fig.~\ref{quantized}-A. For example, assume that there are
two tunnelling junctions with the same resistance and barrier
height, connected in series. In this case, the voltage drop across
each of the junctions would be one-half of the applied voltage,
thus fitting to the I-V curve of a tunnelling junction would yield
a barrier height which would be twice the work function of Au. The
fact that the best fit parameter for the barrier height is only
slightly larger than the work-function of Au indicates that the
lead resistance is much smaller than the resistance $V/I$ of the
junction at $10V$ bias voltage, e.g. $R_{lead}\ll 250k\Omega$ or
$G_{lead}\gg 0.1e^2/h$.

It is striking that tunnelling contacts survive at 10V bias
voltage, since a conventional tunnelling junction with a similar
barrier thickness would typically suffer an electric break down at
10V. We explain the stability of our atomic-scale gaps at 10V with
a dynamical equilibrium between two opposing atom migration
processes.~\cite{korotkov} At 10V, the surface diffusion is
opposed by electromigration (which increases the gap between the
two films).

\subsection{Atomic-Scale Contacts}

After reducing the bias voltage quickly to zero as described
above, we introduce a serial resistor $R_S=20k\Omega$ and start
increasing $U$ at a rate of 1 V/s. The serial resistor limits the
current flowing through the junction, thereby limiting
electromigration. Consequently, the conductance can exceed
$e^2/h$.

At a bias voltage of $U\sim 4V$, conductance of the device begins
to increase in discrete steps as a function of $U$.~\cite{anaya}
An example is shown in Fig.~\ref{quantized}-B. The step size is of
order $0.2 - 2 e^2/h$, suggesting that the junction contains
atomic-scale contacts.  We have recently confirmed these discrete
conductance steps at series resistance $R_S = 100k\Omega$, showing
that the conductance steps are intrinsic to the junction and not
biased by our choice of $R_S$.


In addition to these discrete steps, the conductance changes
continuously as a function of $U$, suggesting that there is a
distributed contribution to the resistance of the junction, from
the leads. In Fig.~\ref{junctions} C and D we show the junctions
from Fig.~\ref{junctions} A and D, respectively, while inside the
SEM, after the conductance was increased to $\approx 2e^2/h$ and
$\approx 6e^2/h$, respectively. One notices that the length of the
junctions increases with conductance. We observe that the
conductance is roughly proportional to the length. The conductance
per unit length is $G/L\approx 600 S/m$. Among different samples,
$G/L$ fluctuates by about a factor of two.

Thus,  the increase in $G$ arises from the addition of Au into the
nano-junction. Notice that the gap in the junction in
Fig.~\ref{junctions}-D remains well defined. We thus arrive at a
model for the nano-junction sketched in Fig.~\ref{model}.
Reservoirs $R_1$ and $R_2$ are bulk Au films, which are good
conductors with sheet resistance of $\approx 5\Omega$. $C_1$ and
$C_2$ are atomic-scale contacts responsible for conductance
quantization. Finally, $L_1$ and $L_2$ are the disordered leads
generated by the atom migration processes. From the images in
Fig.~\ref{junctions}, we obtain that the size of the leads ($D$ in
Fig.~\ref{model}) is approximately 50nm.

Using Ohms laws, the conductance of the junction can be written as
\begin{equation}
G=\sum_i\frac{1}{1/G^i+1/G_{L_1}^i+1/G_{L_1}^i} \label{Ohm}
\end{equation}
where $G^i$ refers to the conductance of an atomic scale contact
in the gap, and $G_{L_{1,2}}^i$ are the conductances between the
contacts and the reservoirs. As the junction dimensions increase,
$G_{L_{1,2}}^i$ changes continuously and $G^i$ changes in discrete
steps of order $e^2/h$.

Because the continuous change in $G$ in Fig.~\ref{quantized}-B is
comparable to the discrete steps in $G$, it follows that the lead
resistance is comparable with the resistance of the atomic-scale
contacts. To obtain the resistivity of the leads, we need to know
the cross-section of these protrusions. Unfortunately we can not
obtain this information through SEM-imaging. If we assume that the
cross-section of the protrusion has the thickness of 50nm, which
is comparable to the film thickness, we obtain $\rho\approx
1.7\cdot 10^5\mu\Omega cm$.

The resistivity is much larger than the maximum metallic
resistivity of $\sim 200\mu\Omega cm$,~\cite{imry} which shows
that the the leads are highly disordered. The disorder is
explained by the intermixing of the impurities into the leads and
grain boundaries.~\cite{anaya} In the device in
Fig.~\ref{junctions}-D (and many other devices), the leads appear
completely uniform down to the imaging resolution (3nm). We still
expect the leads to be granular, with grain diameter (d) smaller
than 3nm, because Au does not form alloys with water (or other
impurities such as $O_2$ and $CO_2$ that are present at $10^{-6}$
Torr background pressure).

In three-dimensional granular systems, the resistance between the
grains ($R_g$) and the resistivity are related as $\rho\sim R_gd$
and it is known that granular systems in 3D exhibit a
metal-insulator transition as a function of $R_g$.~\cite{efetov}
Theoretically, it has been predicted that the transition occurs at
$R_g=R_g^C \sim 19R_Q/ln(E_C/\delta)$, where $E_C$ is the charging
energy of the grain and $\delta$ is the level spacing inside the
grain.~\cite{beloborodov,beloborodov1} In our case, the grain
diameter is less than $3$nm and $\rho\approx 10^5\mu\Omega cm$,
and we estimate $R_g>12h/e^2$ and $R_g^C\sim 5h/e^2$. Thus, we
expect that the electronic states in the leads are strongly
localized.


If the localization length is smaller than the dimensions of the
leads, then the lead resistance at low temperature becomes much
larger than the resistance of the atomic-scale contacts. The
temperature dependence of the resistance becomes significant at
temperatures well below 300K, whereas conductance quantization in
Au is easily observed at room temperature. This explains the
difference between room-temperature and low-temperature properties
of nano-junctions.


\section{Zero Bias Anomalies and Coulomb Blockade\label{cb}}

Electron transport measurements at low temperatures were carried
out using a dilution refrigerator with a base temperature of
$0.015K$. The bias voltage, applied to the sample, was the sum of
a DC-voltage $V$ and an AC voltage with peak-to-peak amplitude
$<10\mu V$ and frequency $<100 Hz$. A current amplifier measured
the current, while lock-in detection from the amplifier output
obtained the differential conductance. The devices were shielded
at $T=0.015K$ by a Faraday cage and home made radiation filters.
The base electron temperature was $\sim 0.05K$.

Transport properties of our junctions changed dramatically when
the temperature was reduced from 300K to 0.015K. At 300K, the
junctions were Ohmic and displayed conductance quantization
effects. At low temperatures, however, the junctions showed
significant suppression near zero-bias voltage.

Devices with $G(300K)<e^2/h$ display Coulomb-blockade at
$T=0.015K$. The Coulomb blockade has been attributed to single
electron charging effects in the grains inside the
leads.~\cite{anaya}

Devices with $G(300K)>2e^2/h$ do not display Coulomb blockade at
$0.015K$. Instead, the conductance versus voltage at $T=0.015K$
displays a ZBA. The ZBAs were interpreted as the Coulomb-Blockade
effect in the strong tunnelling regime.~\cite{anaya}

\subsection{Microscopic Origin of the Charging Effects and ZBAs}

We have found that the Coulomb Blockade in our Au junctions is not
restricted to single-electron charging effects in the grains in
the leads. In fact, the necessary condition to observe
Coulomb-Blockade in our devices is that the leads be highly
resistive, regardless of whether the disorder in the leads is
granular or homogeneous. For example, if we compare
Figs.~\ref{junctions} C and D, we observe that the leads of the
nano-junction in Fig.~\ref{junctions} C have well distinguished
grains, whereas the leads of the nano-junction in
Fig.~\ref{junctions} D are completely uniform. Despite these
differences, the I-V curves of these samples at $T=0.015K$ are
very similar.

We are led to the conclusion that the Coulomb-Blockade and ZBAs at
low temperature arise from localization of electronic states in
the leads, which could either be due to localization of electrons
within one or a few grains (Fig.~\ref{junctions} C), or due to
localization over a region containing a large number of grains
that are too small to observe by the SEM (Fig.~\ref{junctions} D).
In the Coulomb-blockade regime, electron transport is sequential
and takes place via a puddle of electrons in the leads, which is
sketched in Fig.~\ref{model}-B. In sec.~\ref{cf}, we show that the
size of the puddle of electrons in Fig.~\ref{junctions}-D is
comparable to the dimensions of the leads.

Coulomb Blockade in distributed systems has been studied in
disordered $InO_x$ mesoscopic wires.~\cite{chandrasekhar}
Transport properties of these wires exhibited single electron
charging effects at low temperature, very similar to those in
single-electron transistors (SET). However, these wires had no
apparent tunnelling barriers.  The single electron charging
effects were observed if the localization length had been smaller
than the sample size. It was suggested that the size of the
puddles was comparable to the localization length, but it remained
unclear what the junctions were and what formed the puddles of
electrons.

\subsection{Effective Charging Energy}

\begin{figure}
\includegraphics[width=0.75\textwidth]{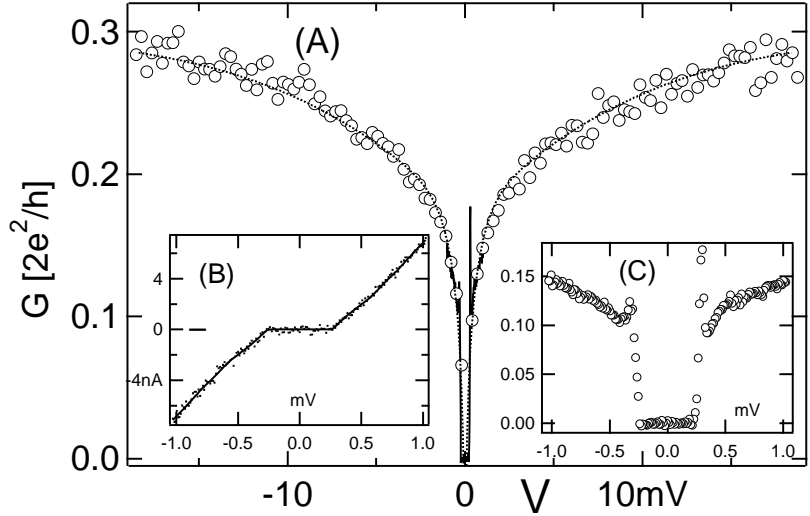}
\caption{A: Differential conductance versus bias voltage of a Au
nano-junction at $T=0.015K$. B: Current versus voltage at
$T=0.015K$ of the nano-junction in a narrow voltage range. C:
Differential conductance versus voltage of the nano-junction at
$T=0.015K$ in a narrow voltage range.~\label{cap}}
\end{figure}

Among devices, the charging energy rapidly decreases as a function
of $G(300K)$. Fig.~\ref{cap}-A shows the conductance versus bias
voltage at $T=0.015K$ in a device with $G(300K)=0.7e^2/h$. This
device belongs to a group of borderline devices in which the
Coulomb blockade is just resolved at $T=0.015K$.

The borderline devices are characterized by two voltage scales. If
the voltage range is large, e.g. $[-20mV,20mV]$ in
Fig.~\ref{cap}-A, the curve resembles ZBAs of high conductance
devices. Thus, in this voltage range we fit the curve to the model
of electron tunnelling through a single-electron transistor in the
strong tunnelling regime.~\cite{golubev} This leads to the
parameter estimates $C_1+C_2=20.8aF$, $R_1=2.7k\Omega$, and
$R_2=34.3k\Omega$, where $C_1$ and $C_2$ are the bare capacitances
between the puddle and the reservoirs, and $R_1$ and $R_2$ are the
bare resistances between the puddle and the reservoirs. The
corresponding bare charging energy is $e^2/2(C_1+C_2)=3.8meV$. The
best fit is shown by the dotted line in Fig.~\ref{cap}-A.

The conductance in Fig.~\ref{cap}-A approaches zero at a nonzero
zero-bias voltage. Fig.~\ref{cap}-C zooms in to Fig.~\ref{cap}-A
around zero bias voltage. The corresponding I-V curve is shown in
Fig.~\ref{cap}-B. The gap in the I-V curve represents
Coulomb-Blockade. By fitting the low bias voltage I-V curves to
the Orthodox theory of single-electron
tunnelling,~\cite{averin,averin2} we estimate $\tilde C_1+\tilde
C_2=442aF$, $\tilde R_1=34k\Omega$, and $\tilde R_2=65k\Omega$ for
the capacitances and the resistances between the puddle and the
reservoirs. The fit is shown by the line in Fig.~\ref{cap}-B. The
corresponding charging energy is $\tilde E_C=e^2/2(\tilde
C_1+\tilde C_2)=0.18meV$, a factor of 21 smaller than the bare
charging energy estimated above.

Theoretically, it has been predicted that Coulomb-Blockade
persists in any diffusive conductor, even if the resistances
between the conductor and the reservoirs is much larger than the
resistance quantum.~\cite{zaikin,nazarov} The persistence of
charging effects in a single electron transistor in strong
coupling to the leads has been demonstrated
experimentally.~\cite{chouvaev,joyez} It has been predicted that
the effective charging energy is given as
\begin{equation}
\tilde E_C=E_Ce^{-\alpha \frac{G}{G_0}}, \label{effective}
\end{equation}
where $G=G_1+G_2$ is the sum of the conductances between the
conductor and the reservoirs, $G_0=2e^2/h$, and, finally, $\alpha$
is a constant of order one.~\cite{zaikin,nazarov}

In our samples, $E_C$ and $\tilde  E_C$ are interpreted as the
bare and effective charging energy, respectively. With
$\alpha\approx 0.6$, they are in rough agreement with
Eq.~\ref{effective}.

\section{Capacitance Fluctuations\label{cf}}

In conventional single-electron transistors, the charging energy
is independent of the applied magnetic field. In  contrast, we
find that the effective charging energy of our nano-junctions
exhibits strong magnetic field dependence.

\begin{figure}
\includegraphics[width=0.75\textwidth]{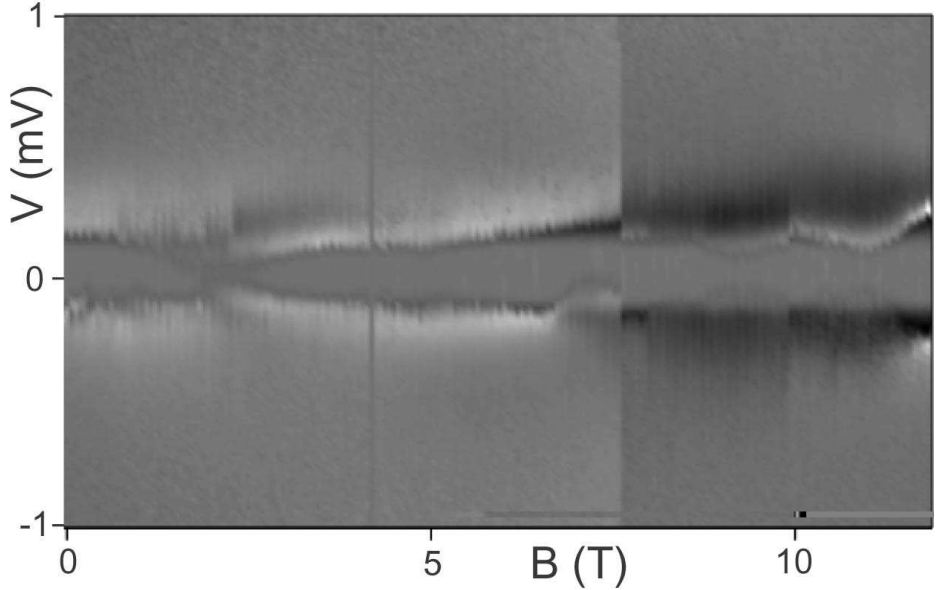}
\caption{Conductance (gray) versus magnetic field and bias voltage
of the nano-junction at $T=0.015K$.~\label{capfl}}
\end{figure}

Fig.~\ref{capfl} displays a gray-scale image of the conductance
versus bias voltage and the applied magnetic field in the sample
with the I-V curves shown in Fig.~\ref{cap}. The magnetic field is
parallel to the slit. The threshold voltage for Coulomb Blockade
(the gap) exhibits a strong non-monotonic dependence -
fluctuations - with the magnetic field. Around the field of 2T,
the gap approaches zero, and around the field of 11T, the gap is
at maximum. The dependence is reproducible when the measurements
are repeated. The amplitude of the gap fluctuations is comparable
to the average gap.

The characteristic magnetic field scale of the gap fluctuations
($B_C$) is given by the typical period of the fluctuations. We
resolve less than a full period in our magnetic field range,
suggesting $B_C\approx 18T$. We have confirmed the gap
fluctuations in 4 additional samples with similar effective
charging energies. The value of $B_C$ is reproducible within a
factor of $2$ among these samples.

\begin{figure}
\includegraphics[width=0.75\textwidth]{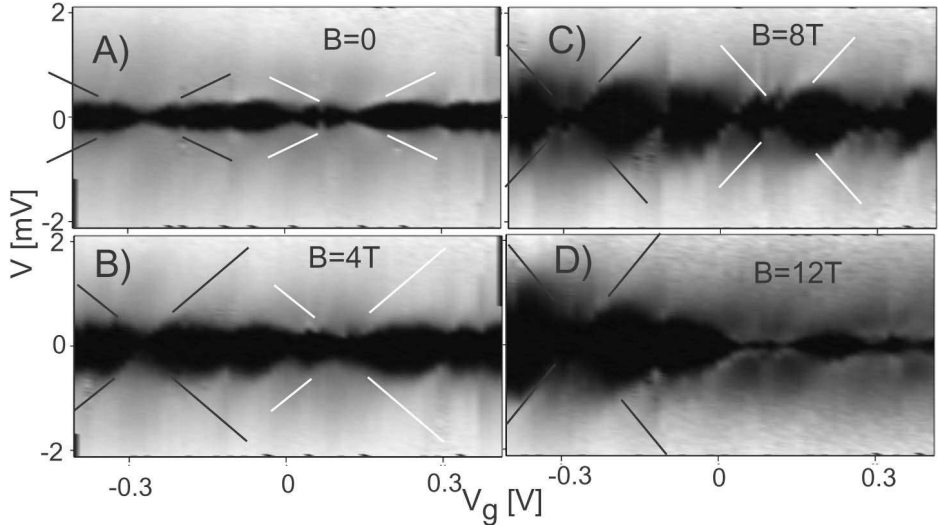}
\caption{A-D: Conductance of a Au nano-junction (gray) versus gate
voltage and bias voltage at four magnetic fields at
$T=0.015K$.~\label{gate}}
\end{figure}

We now show that the fluctuations in the gap represent charging
energy fluctuations (or capacitance fluctuations). To this end, we
examine the gate voltage dependence of the conductance, as a
function of the applied magnetic field. Figure~\ref{gate} displays
conductance versus gate voltage and bias voltage at magnetic
fields of 0T, 4T, 8T, and 12T, in a different sample (the previous
sample did not have a gate). The fabrication of the gate has been
described in Ref.~\cite{anaya}.

Fig.~\ref{gate}-A resembles "diamond diagrams" of conductance
versus gate voltage and bias voltage of quantum dots.~\cite{leo}
The strong dependence of the gap on gate voltage proves that the
gap is caused by the Coulomb-Blockade. In particular, at certain
gate voltages, indicated by the groups of four lines that cross at
a point along the V=0 axis, the gap approaches zero. These points
will be referred to as points where the diamonds close, and the
conductance at these points will refer to the peak conductance
($G_{peak}$). The valley conductance $G_{valley}$ is defined as
the conductance at $V=0$ and at a gate voltage where the gap is at
maximum.

There are significant differences between the diamonds in
Fig.~\ref{gate} and the diamonds of conventional single-electron
transistors. First, the gate voltage dependence of the gap in
Fig.~\ref{gate} is not periodic. We examined the gate voltage
dependence in the range of gate voltages from -2 Volt to 2 Volt,
and found that the structure in Fig.~\ref{cap}-A remained over the
extended voltage range. The structure in Fig.~\ref{cap}-A is
quasiperiodic, in that the slopes of the diamond's edges, near the
points where diamonds close, are the same (i. e. the lines in the
black and the white groups in Fig.~\ref{cap}-A have the same
slopes).

Discontinuities in conductance, as a function of gate voltage,
cause the absence of periodicity in Fig.~\ref{gate}-A. When the
gate voltage sweeps are repeated, conductance discontinuities are
reproducible, and can be attributed to the shifts in the
background charge induced by the changes in gate voltage. The
leads are highly disordered, thus they may contain a large number
of charge traps in the vicinity of the puddle responsible for
Coulomb Blockade. The gate voltage can change the state of the
charge trap, and causes a discontinuous shift in the background
charge.

\begin{figure}
\includegraphics[width=0.75\textwidth]{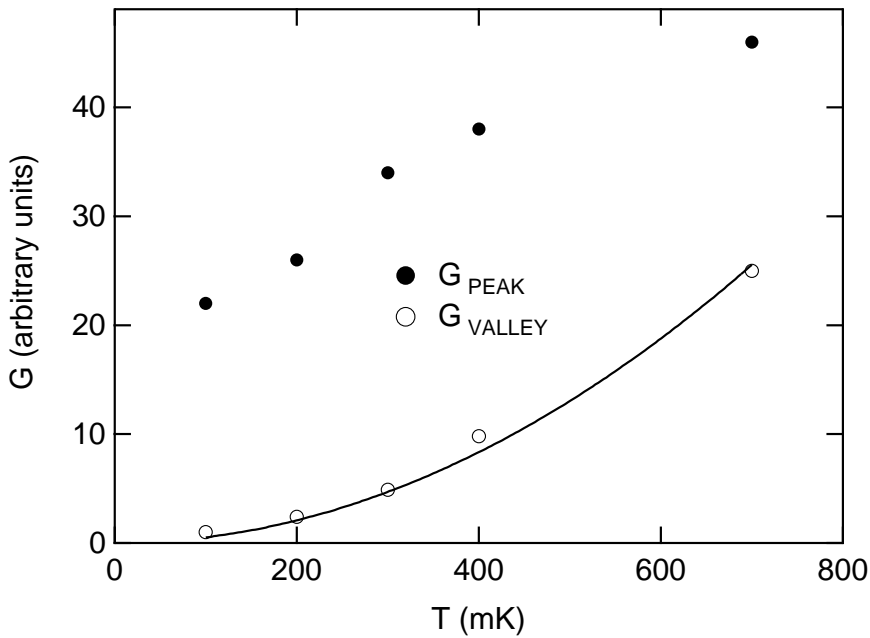}
\caption{Temperature dependence of the peak and the valley
conductance. The line displays the best fit of the valley
conductance to the quadratic temperature dependence.~\label{temp}}
\end{figure}

The second difference between Coulomb Blockade in our
nano-junctions and that of conventional SETs is found in the
conductance peak's temperature dependence. Fig.~\ref{temp} shows
$G_{peak}$ and $G_{valley}$ versus temperature. $G_{peak}$
decreases significantly with temperature even when $k_BT\ll \tilde
E_C$. In contrast, with conventional SETs, $G_{peak}$ has a weak
temperature dependence when $k_BT$ is much smaller than the
charging energy. It appears that $G_{peak}$
 approaches a nonzero
value when $T\to 0$.

At low temperature ($k_BT\ll \tilde E_C$), the valley conductance
goes to zero as $G_{valley}\sim T^2$, as shown in Fig.~\ref{temp}.
The quadratic temperature dependence in the valleys demonstrates
that electron transport in the valleys occurs through inelastic
cotunnelling,~\cite{averin3} which is possible only if the spacing
between energy levels ($\delta$) in the puddle of electrons is
much smaller than $k_BT$. Assuming that the level spacing is given
by $\delta\approx 1/(N(0)V)$, where $N(0)$ is the density of
states at the Fermi level of Au, and $V$ is the volume of the
puddle,  we obtain that $V>(10nm)^3$. This suggests that the
localization length is larger than $10$nm.

At low magnetic fields ($\leq 8T$), we can trace the evolution of
the diamonds with the magnetic field quite well, despite the
discontinuities in the background charge. The points where
diamonds close do not shift with magnetic field in this range.
This implies that the capacitance between the puddle and the gate
($C_g$) does not vary. Therefore, the geometry of the puddle does
not change with magnetic field.

The key effect in Fig.~\ref{gate} A-C is  that it is the puddle's
effective charging energy that changes strongly with magnetic
field. From the Orthodox theory of
Coulomb-Blockade,~\cite{averin2} the slopes of the lines in
Fig.~\ref{gate} are $\pm eC_g/(2\tilde C_1)$ and $\pm
eC_g/(2\tilde C_2)$, where $\tilde C_1$ and $ \tilde C_2$ are the
effective capacitances between the puddle and the reservoirs. It
follows that $\tilde C_1$ and $\tilde C_2$ fluctuate with field.
In particular, from Fig.~\ref{gate}, we obtain $\tilde
C_1(4T)=1.8\tilde C_1(0)$, $\tilde C_1(8T)=2.3\tilde C_1(0)$, and
$\tilde C_1(12T)=2.4\tilde C_1(0)$.

Nazarov had predicted fluctuations of effective capacitance in
coherent conductors in the regime of strong coupling to the
reservoirs.~\cite{nazarov} With strong coupling, the
Coulomb-Blockade survives in any coherent disordered conductor. In
this regime, effective capacitance exhibits mesoscopic
fluctuations as a function of the applied magnetic field. These
fluctuations are analogous to universal conductance
fluctuations.~\cite{washburn}

One way to understand capacitance fluctuations is to observe that
the effective charging energy, $e^2/2(\tilde C_1+\tilde C_2$),
exponentially depends  on the conductance between the conductor
and the reservoirs, see Eq.~\ref{effective}. Then, the universal
conductance fluctuations induce fluctuations in $G_1+G_2$ with
field, which leads to the fluctuations in the effective charging
energy. Since the amplitude of conductance fluctuations in the
diffusive regime is $\sim G_Q$, it follows that the amplitude of
the charging energy fluctuations is comparable to the average
charging energy, consistent with our data.

We expect that the characteristic magnetic field scale is given by
the flux quantum ($\Phi_0=h/2e$) over the directed area of the
puddle, $B_C\sim \Phi_0/L_s^2$, where $L_S$ is the diameter of the
puddle (localization length). For $B_C\approx 18T$, we obtain
$L_s\approx 105$nm, which is comparable to the dimensions of the
leads $D\approx 50$nm.

If the magnetic field approaches 12T, it becomes hard to trace the
diamonds. In fact, at the field of $12T$ (Fig.~\ref{gate}-D), the
structure is no longer quasiperiodic. This suggests that when the
magnetic field approaches $B_C$, conduction can no longer be
described by sequential tunnelling via the same puddle of
electrons. The strong-field regime is the subject of current
research.

\section{Comparison with Electroplated Nano-junctions\label{electro}}

Our introduction described strong ZBAs, observed in electroplated
Au nano-junctions containing atomic-scale contacts.~\cite{yu,yu1}
The bulk electroplated material was found not to undergo a strong
localization transition at low temperatures. In addition, the ZBAs
exhibited scaling with junction size that could not be easily
explained in the localization framework. The scaling suggested
that the ZBAs displayed a suppression in the density of states in
the leads.

Note that only if the resistance of atomic-scale contacts is much
larger than the lead resistance can the conductance of the contact
be proportional to the density of states in the leads, as would be
the case in conventional tunnelling junctions.~\cite{altshuler}
Thus, a $\sim 100$\% suppression of the density of states in
electroplated nano-junctions must be very local around the
atomic-scale contact. If it were otherwise, the conductivity of
the leads would be $\sim 100$\% reduced in a region much larger
than the contact size, and the lead resistance would not be much
smaller than the resistance of the contact.

In our devices, the localization length is comparable to the
dimensions of the leads and the conductance is not proportional to
the density of states. The ZBAs in our nano-junctions are caused
by the Coulomb blockade on localized puddles of electrons inside
the leads, analogous to Coulomb-Blockade in disordered $InO_x$
wires.~\cite{chandrasekhar} The ZBAs are manifestations of
Coulomb-Blockade on these puddles in the regime of strong-coupling
to the reservoirs.

\section{Concluding Remarks}

Atomic-scale point contacts of Au between strongly disordered
leads can have striking differences between their room-temperature
and the low temperature properties. At room temperature the
contacts exhibit conductance quantization and are Ohmic, at low
temperatures the contacts exhibit Coulomb-Blockade or zero-bias
anomalies. The differences between the room-temperature and the
low temperature properties arise from the localization of
electronic states in the leads. The temperature at which the
resistance of the leads becomes significantly larger than the
resistance of the contacts is much lower than the room
temperature.

At low temperature, Coulomb-Blockade arises from puddles of
electrons in the leads that form as a result of localization. One
can distinguish between the bare charging energy and the effective
charging energy of the puddles. The latter is found to be much
smaller than the former, and it exhibits strong fluctuations with
an applied magnetic field. The gate voltage effects of a magnetic
field demonstrate that the effective capacitance between the
puddle and the reservoirs fluctuates with the magnetic field, in
agreement with theoretical predictions.

\begin{acknowledgments}
This work was performed in part at the Cornell Nanofabrication
Facility, (a member of the National Nanofabrication Users
Network), which is supported by the NSF, under grant ECS-9731293,
Cornell University and Industrial affiliates, and the Georgia-Tech
electron microscopy facility. This research is supported by the
David and Lucile Packard Foundation grant 2000-13874 and the NSF
grant DMR-0102960.
\end{acknowledgments}

\bibliography{nanotech}

\begin{thebibliography}{28}
\expandafter\ifx\csname natexlab\endcsname\relax\def\natexlab#1{#1}\fi
\expandafter\ifx\csname bibnamefont\endcsname\relax
  \def\bibnamefont#1{#1}\fi
\expandafter\ifx\csname bibfnamefont\endcsname\relax
  \def\bibfnamefont#1{#1}\fi
\expandafter\ifx\csname citenamefont\endcsname\relax
  \def\citenamefont#1{#1}\fi
\expandafter\ifx\csname url\endcsname\relax
  \def\url#1{\texttt{#1}}\fi
\expandafter\ifx\csname urlprefix\endcsname\relax\def\urlprefix{URL }\fi
\providecommand{\bibinfo}[2]{#2}
\providecommand{\eprint}[2][]{\url{#2}}

\bibitem[{\citenamefont{Muller et~al.}(1992)\citenamefont{Muller, van
  Ruitenbeek, and de~Jongh}}]{muller}
\bibinfo{author}{\bibfnamefont{C.~J.} \bibnamefont{Muller}},
  \bibinfo{author}{\bibfnamefont{J.~M.} \bibnamefont{van Ruitenbeek}},
  \bibnamefont{and} \bibinfo{author}{\bibfnamefont{L.~J.}
  \bibnamefont{de~Jongh}}, \bibinfo{journal}{Physica C}
  \textbf{\bibinfo{volume}{191}}, \bibinfo{pages}{485} (\bibinfo{year}{1992}).

\bibitem[{\citenamefont{Agrait et~al.}(2003)\citenamefont{Agrait, Yeyati, and
  Ruitenbeek}}]{agrait}
\bibinfo{author}{\bibfnamefont{N.}~\bibnamefont{Agrait}},
  \bibinfo{author}{\bibfnamefont{A.~L.} \bibnamefont{Yeyati}},
  \bibnamefont{and} \bibinfo{author}{\bibfnamefont{J.~M.~V.}
  \bibnamefont{Ruitenbeek}}, \bibinfo{journal}{Phys. Rep.}
  \textbf{\bibinfo{volume}{377}}, \bibinfo{pages}{81} (\bibinfo{year}{2003}).

\bibitem[{\citenamefont{Morpurgo et~al.}(1999)\citenamefont{Morpurgo, Marcus,
  and Robinson}}]{morpurgo}
\bibinfo{author}{\bibfnamefont{A.~F.} \bibnamefont{Morpurgo}},
  \bibinfo{author}{\bibfnamefont{C.~M.} \bibnamefont{Marcus}},
  \bibnamefont{and} \bibinfo{author}{\bibfnamefont{D.~B.}
  \bibnamefont{Robinson}}, \bibinfo{journal}{Appl Phys Lett}
  \textbf{\bibinfo{volume}{74}}, \bibinfo{pages}{2084} (\bibinfo{year}{1999}).

\bibitem[{\citenamefont{Li et~al.}(2000)\citenamefont{Li, He, A, Bunch, and
  Tao}}]{li}
\bibinfo{author}{\bibfnamefont{C.~Z.} \bibnamefont{Li}},
  \bibinfo{author}{\bibfnamefont{H.~X.} \bibnamefont{He}},
  \bibinfo{author}{\bibfnamefont{A.~B.} \bibnamefont{A}},
  \bibinfo{author}{\bibfnamefont{J.~S.} \bibnamefont{Bunch}}, \bibnamefont{and}
  \bibinfo{author}{\bibfnamefont{N.~J.} \bibnamefont{Tao}},
  \bibinfo{journal}{Phys. Rev. A} \textbf{\bibinfo{volume}{76}},
  \bibinfo{pages}{1333} (\bibinfo{year}{2000}).

\bibitem[{\citenamefont{Boussaad and Tao}(2002)}]{boussaad}
\bibinfo{author}{\bibfnamefont{S.}~\bibnamefont{Boussaad}} \bibnamefont{and}
  \bibinfo{author}{\bibfnamefont{N.~J.} \bibnamefont{Tao}},
  \bibinfo{journal}{Appl Phys Lett} \textbf{\bibinfo{volume}{80}},
  \bibinfo{pages}{2398} (\bibinfo{year}{2002}).

\bibitem[{\citenamefont{Yu and Natelson}(2003{\natexlab{a}})}]{yu}
\bibinfo{author}{\bibfnamefont{L.~H.} \bibnamefont{Yu}} \bibnamefont{and}
  \bibinfo{author}{\bibfnamefont{D.}~\bibnamefont{Natelson}},
  \bibinfo{journal}{Appl Phys Lett} \textbf{\bibinfo{volume}{82}},
  \bibinfo{pages}{2332} (\bibinfo{year}{2003}{\natexlab{a}}).

\bibitem[{\citenamefont{Yu and Natelson}(2003{\natexlab{b}})}]{yu1}
\bibinfo{author}{\bibfnamefont{L.~H.} \bibnamefont{Yu}} \bibnamefont{and}
  \bibinfo{author}{\bibfnamefont{D.}~\bibnamefont{Natelson}},
  \bibinfo{journal}{Phys. Rev. B} \textbf{\bibinfo{volume}{68}},
  \bibinfo{pages}{113407} (\bibinfo{year}{2003}{\natexlab{b}}).

\bibitem[{\citenamefont{Park et~al.}(1999)\citenamefont{Park, Lim, Alivisatos,
  Park, and McEuen}}]{park}
\bibinfo{author}{\bibfnamefont{H.}~\bibnamefont{Park}},
  \bibinfo{author}{\bibfnamefont{A.~K.~L.} \bibnamefont{Lim}},
  \bibinfo{author}{\bibfnamefont{A.~P.} \bibnamefont{Alivisatos}},
  \bibinfo{author}{\bibfnamefont{J.}~\bibnamefont{Park}}, \bibnamefont{and}
  \bibinfo{author}{\bibfnamefont{P.~L.} \bibnamefont{McEuen}},
  \bibinfo{journal}{Appl. Phys. Lett.} \textbf{\bibinfo{volume}{75}},
  \bibinfo{pages}{301} (\bibinfo{year}{1999}).

\bibitem[{\citenamefont{Park et~al.}(2002)\citenamefont{Park, Pasupathy,
  Goldsmith, Chang, Yaish, Petta, Rinkoski, Sethna, Abruna, McEuen
  et~al.}}]{park1}
\bibinfo{author}{\bibfnamefont{J.}~\bibnamefont{Park}},
  \bibinfo{author}{\bibfnamefont{A.~N.} \bibnamefont{Pasupathy}},
  \bibinfo{author}{\bibfnamefont{J.~I.} \bibnamefont{Goldsmith}},
  \bibinfo{author}{\bibfnamefont{C.}~\bibnamefont{Chang}},
  \bibinfo{author}{\bibfnamefont{Y.}~\bibnamefont{Yaish}},
  \bibinfo{author}{\bibfnamefont{J.~R.} \bibnamefont{Petta}},
  \bibinfo{author}{\bibfnamefont{M.}~\bibnamefont{Rinkoski}},
  \bibinfo{author}{\bibfnamefont{J.~P.} \bibnamefont{Sethna}},
  \bibinfo{author}{\bibfnamefont{H.~D.} \bibnamefont{Abruna}},
  \bibinfo{author}{\bibfnamefont{P.~L.} \bibnamefont{McEuen}},
  \bibnamefont{et~al.}, \bibinfo{journal}{Nature}
  \textbf{\bibinfo{volume}{417}}, \bibinfo{pages}{722} (\bibinfo{year}{2002}).

\bibitem[{\citenamefont{Anaya et~al.}(2003)\citenamefont{Anaya, Korotkov,
  Bowman, Waddell, and Davidovi\'c}}]{anaya}
\bibinfo{author}{\bibfnamefont{A.}~\bibnamefont{Anaya}},
  \bibinfo{author}{\bibfnamefont{A.~L.} \bibnamefont{Korotkov}},
  \bibinfo{author}{\bibfnamefont{M.}~\bibnamefont{Bowman}},
  \bibinfo{author}{\bibfnamefont{J.}~\bibnamefont{Waddell}}, \bibnamefont{and}
  \bibinfo{author}{\bibfnamefont{D.}~\bibnamefont{Davidovi\'c}},
  \bibinfo{journal}{Journal of Applied Physics} \textbf{\bibinfo{volume}{93}},
  \bibinfo{pages}{3501} (\bibinfo{year}{2003}).

\bibitem[{\citenamefont{Averin and Korotkov}(1990)}]{averin}
\bibinfo{author}{\bibfnamefont{D.~V.} \bibnamefont{Averin}} \bibnamefont{and}
  \bibinfo{author}{\bibfnamefont{A.~N.} \bibnamefont{Korotkov}},
  \bibinfo{journal}{Journal of low temperature physics}
  \textbf{\bibinfo{volume}{80}}, \bibinfo{pages}{173} (\bibinfo{year}{1990}).

\bibitem[{\citenamefont{Averin and Likharev}(1991)}]{averin2}
\bibinfo{author}{\bibfnamefont{D.~V.} \bibnamefont{Averin}} \bibnamefont{and}
  \bibinfo{author}{\bibfnamefont{K.~K.} \bibnamefont{Likharev}}, in
  \emph{\bibinfo{booktitle}{Mesoscopic Phenomena in Solids}}, edited by
  \bibinfo{editor}{\bibfnamefont{B.~L.} \bibnamefont{Altshuler}},
  \bibinfo{editor}{\bibfnamefont{P.~L.} \bibnamefont{Lee}}, \bibnamefont{and}
  \bibinfo{editor}{\bibfnamefont{R.~A.} \bibnamefont{Webb}}
  (\bibinfo{publisher}{Elsevier and Amsterdam}, \bibinfo{year}{1991}), p.
  \bibinfo{pages}{169}.

\bibitem[{\citenamefont{Golubev et~al.}(1997)\citenamefont{Golubev, Konig,
  Schoeller, Schon, and Zaikin}}]{golubev}
\bibinfo{author}{\bibfnamefont{D.~S.} \bibnamefont{Golubev}},
  \bibinfo{author}{\bibfnamefont{J.}~\bibnamefont{Konig}},
  \bibinfo{author}{\bibfnamefont{H.}~\bibnamefont{Schoeller}},
  \bibinfo{author}{\bibfnamefont{G.}~\bibnamefont{Schon}}, \bibnamefont{and}
  \bibinfo{author}{\bibfnamefont{A.~D.} \bibnamefont{Zaikin}},
  \bibinfo{journal}{Phys. Rev. B} \textbf{\bibinfo{volume}{56}},
  \bibinfo{pages}{15782} (\bibinfo{year}{1997}).

\bibitem[{\citenamefont{Nazarov}(1999)}]{nazarov}
\bibinfo{author}{\bibfnamefont{Y.~V.} \bibnamefont{Nazarov}},
  \bibinfo{journal}{Phys. Rev. Lett.} \textbf{\bibinfo{volume}{82}},
  \bibinfo{pages}{1245} (\bibinfo{year}{1999}).

\bibitem[{\citenamefont{Korotkov et~al.}(2003)\citenamefont{Korotkov, Bowman,
  McGuinness, and Davidovic}}]{korotkov}
\bibinfo{author}{\bibfnamefont{A.~L.} \bibnamefont{Korotkov}},
  \bibinfo{author}{\bibfnamefont{M.}~\bibnamefont{Bowman}},
  \bibinfo{author}{\bibfnamefont{H.~J.} \bibnamefont{McGuinness}},
  \bibnamefont{and}
  \bibinfo{author}{\bibfnamefont{D.}~\bibnamefont{Davidovic}},
  \bibinfo{journal}{Nanotechnology} \textbf{\bibinfo{volume}{14}},
  \bibinfo{pages}{42} (\bibinfo{year}{2003}).

\bibitem[{\citenamefont{T.M.Mayer et~al.}(1999)\citenamefont{T.M.Mayer,
  J.E.Hueston, G.E.Franklin, A.A.Erchak, and Michalske}}]{mayer}
\bibinfo{author}{\bibnamefont{T.M.Mayer}},
  \bibinfo{author}{\bibnamefont{J.E.Hueston}},
  \bibinfo{author}{\bibnamefont{G.E.Franklin}},
  \bibinfo{author}{\bibnamefont{A.A.Erchak}}, \bibnamefont{and}
  \bibinfo{author}{\bibnamefont{Michalske}}, \bibinfo{journal}{Jour. Appl.
  Phys.} \textbf{\bibinfo{volume}{85}}, \bibinfo{pages}{8170}
  (\bibinfo{year}{1999}).

\bibitem[{\citenamefont{Imry}(1997)}]{imry}
\bibinfo{author}{\bibfnamefont{Y.}~\bibnamefont{Imry}},
  \emph{\bibinfo{title}{Introduction to mesoscopic physics}}
  (\bibinfo{publisher}{Oxford University Press}, \bibinfo{year}{1997}).

\bibitem[{\citenamefont{Efetov and Tschersich}(2003)}]{efetov}
\bibinfo{author}{\bibfnamefont{K.~B.} \bibnamefont{Efetov}} \bibnamefont{and}
  \bibinfo{author}{\bibfnamefont{A.}~\bibnamefont{Tschersich}},
  \bibinfo{journal}{Phys. Rev. B} \textbf{\bibinfo{volume}{67}},
  \bibinfo{pages}{174205} (\bibinfo{year}{2003}).

\bibitem[{\citenamefont{Beloborodov et~al.}(2001)\citenamefont{Beloborodov,
  Efetov, Altland, and Hekking}}]{beloborodov}
\bibinfo{author}{\bibfnamefont{I.~S.} \bibnamefont{Beloborodov}},
  \bibinfo{author}{\bibfnamefont{K.~B.} \bibnamefont{Efetov}},
  \bibinfo{author}{\bibfnamefont{A.}~\bibnamefont{Altland}}, \bibnamefont{and}
  \bibinfo{author}{\bibfnamefont{F.~W.~J.} \bibnamefont{Hekking}},
  \bibinfo{journal}{Phys. Rev. B} \textbf{\bibinfo{volume}{63}},
  \bibinfo{pages}{115109} (\bibinfo{year}{2001}).

\bibitem[{\citenamefont{Beloborodov et~al.}()\citenamefont{Beloborodov, Efetov,
  Lopatin, and Vinokur}}]{beloborodov1}
\bibinfo{author}{\bibfnamefont{I.~S.} \bibnamefont{Beloborodov}},
  \bibinfo{author}{\bibfnamefont{K.~B.} \bibnamefont{Efetov}},
  \bibinfo{author}{\bibfnamefont{A.~V.} \bibnamefont{Lopatin}},
  \bibnamefont{and} \bibinfo{author}{\bibfnamefont{V.~M.}
  \bibnamefont{Vinokur}},
  \bibinfo{journal}{http://xxx.lanl.gov/abs/cond-mat/0304448}  (????).

\bibitem[{\citenamefont{Chandrasekhar et~al.}(1991)\citenamefont{Chandrasekhar,
  Ovadyahu, and Webb}}]{chandrasekhar}
\bibinfo{author}{\bibfnamefont{V.}~\bibnamefont{Chandrasekhar}},
  \bibinfo{author}{\bibfnamefont{Z.}~\bibnamefont{Ovadyahu}}, \bibnamefont{and}
  \bibinfo{author}{\bibfnamefont{R.~A.} \bibnamefont{Webb}},
  \bibinfo{journal}{Phys. Rev. Lett} \textbf{\bibinfo{volume}{67}},
  \bibinfo{pages}{2862} (\bibinfo{year}{1991}).

\bibitem[{\citenamefont{Panyukov and Zaikin}(1991)}]{zaikin}
\bibinfo{author}{\bibfnamefont{S.~V.} \bibnamefont{Panyukov}} \bibnamefont{and}
  \bibinfo{author}{\bibfnamefont{A.~D.} \bibnamefont{Zaikin}},
  \bibinfo{journal}{Phys. Rev. Lett.} \textbf{\bibinfo{volume}{67}},
  \bibinfo{pages}{3168} (\bibinfo{year}{1991}).

\bibitem[{\citenamefont{Chouvaev et~al.}(1999)\citenamefont{Chouvaev, Kuzmin,
  Golubev, and Zaikin}}]{chouvaev}
\bibinfo{author}{\bibfnamefont{D.}~\bibnamefont{Chouvaev}},
  \bibinfo{author}{\bibfnamefont{L.~S.} \bibnamefont{Kuzmin}},
  \bibinfo{author}{\bibfnamefont{D.~S.} \bibnamefont{Golubev}},
  \bibnamefont{and} \bibinfo{author}{\bibfnamefont{A.~D.}
  \bibnamefont{Zaikin}}, \bibinfo{journal}{Phys. Rev. B}
  \textbf{\bibinfo{volume}{59}}, \bibinfo{pages}{10599} (\bibinfo{year}{1999}).

\bibitem[{\citenamefont{Joyez et~al.}(1997)\citenamefont{Joyez, Bouchiat,
  Esteve, Urbina, and Devoret}}]{joyez}
\bibinfo{author}{\bibfnamefont{P.}~\bibnamefont{Joyez}},
  \bibinfo{author}{\bibfnamefont{V.}~\bibnamefont{Bouchiat}},
  \bibinfo{author}{\bibfnamefont{D.}~\bibnamefont{Esteve}},
  \bibinfo{author}{\bibfnamefont{C.}~\bibnamefont{Urbina}}, \bibnamefont{and}
  \bibinfo{author}{\bibfnamefont{M.~H.} \bibnamefont{Devoret}},
  \bibinfo{journal}{Phys. Rev. Lett.} \textbf{\bibinfo{volume}{79}},
  \bibinfo{pages}{1349} (\bibinfo{year}{1997}).

\bibitem[{\citenamefont{Kouwenhoven et~al.}(1997)\citenamefont{Kouwenhoven,
  Marcus, McEuen, Tarucha, Westervelt, and Wingreen}}]{leo}
\bibinfo{author}{\bibfnamefont{L.~P.} \bibnamefont{Kouwenhoven}},
  \bibinfo{author}{\bibfnamefont{C.}~\bibnamefont{Marcus}},
  \bibinfo{author}{\bibfnamefont{P.}~\bibnamefont{McEuen}},
  \bibinfo{author}{\bibfnamefont{S.}~\bibnamefont{Tarucha}},
  \bibinfo{author}{\bibfnamefont{R.}~\bibnamefont{Westervelt}},
  \bibnamefont{and} \bibinfo{author}{\bibfnamefont{N.}~\bibnamefont{Wingreen}},
  in \emph{\bibinfo{booktitle}{Mesoscopic Electron Transport}}, edited by
  \bibinfo{editor}{\bibfnamefont{L.~P.} \bibnamefont{Kouwenhoven}},
  \bibinfo{editor}{\bibfnamefont{L.~L.} \bibnamefont{Sohn}}, \bibnamefont{and}
  \bibinfo{editor}{\bibfnamefont{G.}~\bibnamefont{Schon}}
  (\bibinfo{publisher}{Elsevier and Amsterdam}, \bibinfo{year}{1997}), p.
  \bibinfo{pages}{549}.

\bibitem[{\citenamefont{Averin and Nazarov}(1991)}]{averin3}
\bibinfo{author}{\bibfnamefont{D.~V.} \bibnamefont{Averin}} \bibnamefont{and}
  \bibinfo{author}{\bibfnamefont{Y.~V.} \bibnamefont{Nazarov}}
  (\bibinfo{publisher}{Kluwer Academic/Plenum Publishers},
  \bibinfo{year}{1991}).

\bibitem[{\citenamefont{Washburn and Webb}(1992)}]{washburn}
\bibinfo{author}{\bibfnamefont{S.}~\bibnamefont{Washburn}} \bibnamefont{and}
  \bibinfo{author}{\bibfnamefont{R.~A.} \bibnamefont{Webb}},
  \bibinfo{journal}{Rep. Prog. Phys.} \textbf{\bibinfo{volume}{55}},
  \bibinfo{pages}{1311} (\bibinfo{year}{1992}).

\bibitem[{\citenamefont{Altshuler and Aronov}(1985)}]{altshuler}
\bibinfo{author}{\bibfnamefont{B.~L.} \bibnamefont{Altshuler}}
  \bibnamefont{and} \bibinfo{author}{\bibfnamefont{A.~G.}
  \bibnamefont{Aronov}}, in \emph{\bibinfo{booktitle}{Electron-Electron
  Interactions in Disordered Systems}}, edited by
  \bibinfo{editor}{\bibfnamefont{A.~L.} \bibnamefont{Efros}} \bibnamefont{and}
  \bibinfo{editor}{\bibfnamefont{M.}~\bibnamefont{Pollak}}
  (\bibinfo{publisher}{Elsevier and Amsterdam}, \bibinfo{year}{1985}).

\end{thebibliography}

\end{document}